\documentclass[twocolumn]{article}
\usepackage{amsmath}       
\usepackage{graphicx}      
\usepackage{epstopdf}      
\usepackage{booktabs}      
\usepackage{siunitx}       
\usepackage{caption}       
\usepackage{url}           
\usepackage{abstract}		
\newcommand{\figref}[2][\figurename~]{#1\ref{#2}}

\usepackage[backend=biber,style=nature]{biblatex} 
\begin{document}
\title{Switchable selective backscatter modulation via the Kerker effect} 
\author{Kieran J. Cowan,$^{1, a)}$ Calvin M. Hooper,$^1$ Cameron P. Gallagher,$^1$ Simon J. Berry,$^2$ \and Alastair P. Hibbins,$^1$ and Alexander W. Powell,$^1$}

\date{$^1$Department of Physics and Astronomy, Centre for Metamaterial Research and Innovation, University of Exeter, Stocker Road, Exeter, EX4 4QL, UK\\
    $^2$QinetiQ, Farnborough, Hampshire, GU14 0LX, UK\\
    $^{a)}$kc485@exeter.ac.uk\\[2ex]}
    
\twocolumn[	
\maketitle 
\begin{onecolabstract}
\noindent
We design a reflective scatterer that operates only in the backscatter direction using modulation of the Kerker effect. This being a phenomena where the resonant modes of a scattering object produce total destructive interference in the backscatter direction. Our design consists of a dielectric shell and an interior metal element that provide the required dipolar contributions for Kerker interference with angular stability. By incorporating a PIN diode, we obtain a scatterer that exhibits strong amplitude modulation in the backscattering direction that is weaker at other bistatic angles. Backscatter modulation is a key technology to connect everyday objects but can be vulnerable to eavesdropping, the device presented in this work may help guard against such flaws.

\end{onecolabstract}
\vspace{5 mm} 
]

\section{Introduction}
\label{sec:introduction}
Utilising the far field interference between different resonant modes to control the directional scattering of an object has a long history across fields from radio frequency antennas to nano-optics \cite{Yagi1928,LiuMiroshnichenk2018}. One such phenomena is the Kerker effect, where certain combinations of the multipole resonances of a scattering object, such as an electric and magnetic dipole, produce complete destructive interference in the backward scattering \cite{W.Liu2018}. The destructive interference of the backscatter has naturally led to the Kerker effect being explored in the ongoing development of reflectionless surfaces \cite{Sharma2023, Londono2018, W.Liu2018}. Conversely, by superimposing resonances in this manner it is possible to obtain constructive interference that is of interest in beam steering and superdirective antennas \cite{Ataloglou2022, Gaponenko2023, M.Liu2018, W.Liu2018, Zhang2024, Wu2020}.

This approach has been applied across several fields within electromagnetism and beyond. At the nanoscale, optical Kerker effect modulation has been theoretically proposed via the switching of applied magnetic fields and different excitation methods of the dipoles and has experimentally been achieved using temperature variation \cite{Zouros2020, Wei2017, Oguntoye2023ContinuouslyMetasurfaces}. At microwave frequencies, Kerker effect modulation has been experimentally realised using diodes in the unit cells of arrays and on the elements of antennas \cite{Ataloglou2022, M.Liu2018, Wu2020, Duan2025}. It has been used to change the sideband scattering of a 1D Huygens' source array, to switch endfire direction of an antenna and to shift the direction of destructive interference from an antenna in four axis directions \cite{M.Liu2018, Wu2020, Duan2025}. However, the modulation of a single reflective scatterer in order to switch the Kerker effect `on' and `off' has yet to be experimentally realised. This is despite the recent interest in developing more secure backscatter communications, to protect signals encoded via small scatterers against unwanted observers  \cite{Kumar2021,Lu2024, Wang2021}. A scatterer that only exhibited modulation in the backscatter direction would be more secure than, for example, a simple dipole scatterer whose switching on and off is observable at any angle in a plane. Directional modulation achieves this by obscuring the true modulation except in a target direction \cite{Ansari2022}. Previous works have explored various approaches to achieve this, but in general these require multiple antenna elements all independently controlled, potentially making them too large and complex for many applications \cite{Babakhani2008, Yang2016, Wang2021, Huang2022, Ansari2022}. To our knowledge the possible use of a switchable Kerker effect for increasing the security of backscatter communications has not been explored.

\begin{figure*}[h!]{}
\centering
\includegraphics[width=160mm]{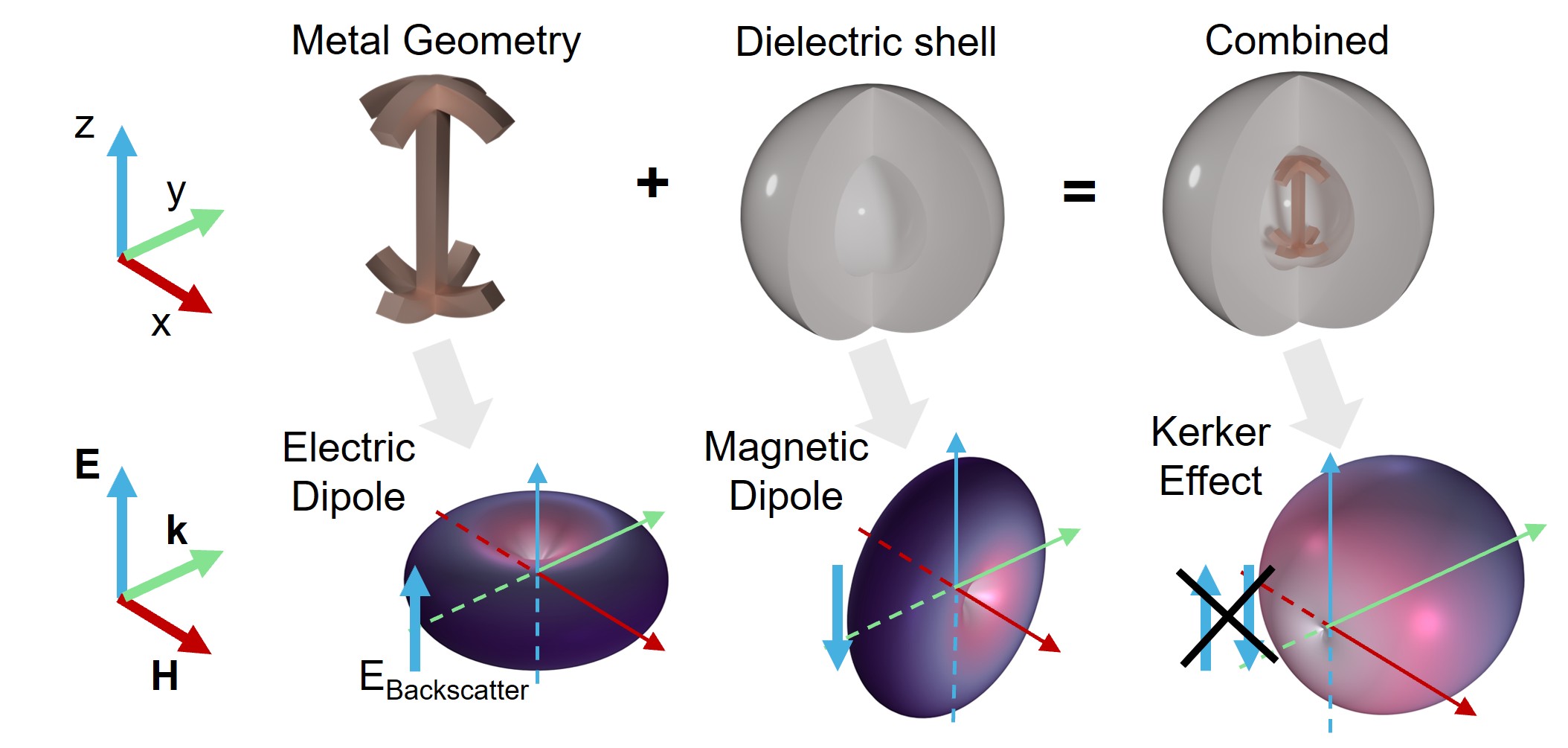}
\caption{Schematic showing the components of our design and the radiation patterns of their contributions towards obtaining the Kerker effect for a wave incident in the x-y plane with electric field polarisation aligned to the z-axis. In the Kerker effect the electric far field in the backscatter direction of an electric and magnetic dipole cancel out, giving a null. A section of the shell is missing in this visual representation so that the core is visible.}
\label{fig:visual}
\end{figure*}

In this paper we design a scatterer that exhibits destructive interference only in the backscatter direction (a Kerker effect minimum) that can be significantly modulated whilst contrast at other scattering angles is weaker. This effect holds true for any incident angle in the horizontal plane. The switching is achieved using a PIN diode at the centre of a metal element that provides primarily an electric dipole contribution, allowing it to be effectively turned `off'. This metal element is located within the core of a dielectric shell that is providing an orthogonal magnetic dipole. The two contributions interfere to result in a minimum in the backscatter radiation as explained schematically in \figref{fig:visual}. In addition to experimentally switching our structure we show that, in a plane perpendicular to the electric dipole, the Kerker effect is obtained from any direction and that at angles other than backscatter the amplitude contrast is significantly reduced. This work could form the foundation of a new type of directionally secure scatterer for backscatter communications.

\section{Design}
\label{sec:background}
Kerker et al. discovered that in the multipole expansion of the modes of a sphere, specific conditions can lead to the total suppression of the backscattering component \cite{Kerker1983ElectromagneticSpheres}. This scattering concept has since been extended to include non-spherical structures and termed the Kerker effect \cite{W.Liu2018,Zouros2020,Wei2017, M.Liu2018, Jing2025, Wen2025, Igoshin2025}. Considering only dipolar contributions, the Kerker effect can occur when an electric and an orthogonal magnetic dipole are co-located with equal magnitudes and phase. The dipole moments need to be orthogonal to each other so that in the backscatter direction their electric far fields will be aligned but out-of-phase and so destructively interfere (\figref{fig:visual}). When the Kerker effect occurs only due to dipole contributions, as opposed to including higher order multipoles, it is sometimes referred to as a Huygens' source \cite{Kozlov2016, Sharma2023, M.Liu2018, Zhang2024, Wu2020}.

Previous work has examined manipulating multipoles in simple dielectric spheres \cite{Powell2021,Liu2014}, demonstrating orthogonal electric and magnetic dipoles that give a minimum in backscatter radiation due to the Kerker effect. The work showed that the introduction of a metal core to the dielectric sphere changes the overall electric dipole moment, thus creating a second minimum and shifting the frequency of the first minimum \cite{Powell2021}. For this symmetrical system, the minima are in the backscattering direction for any angle of the incident wave due to the spherical symmetry. 

\begin{figure}[h!]{}
\centering
\includegraphics[width=80mm]{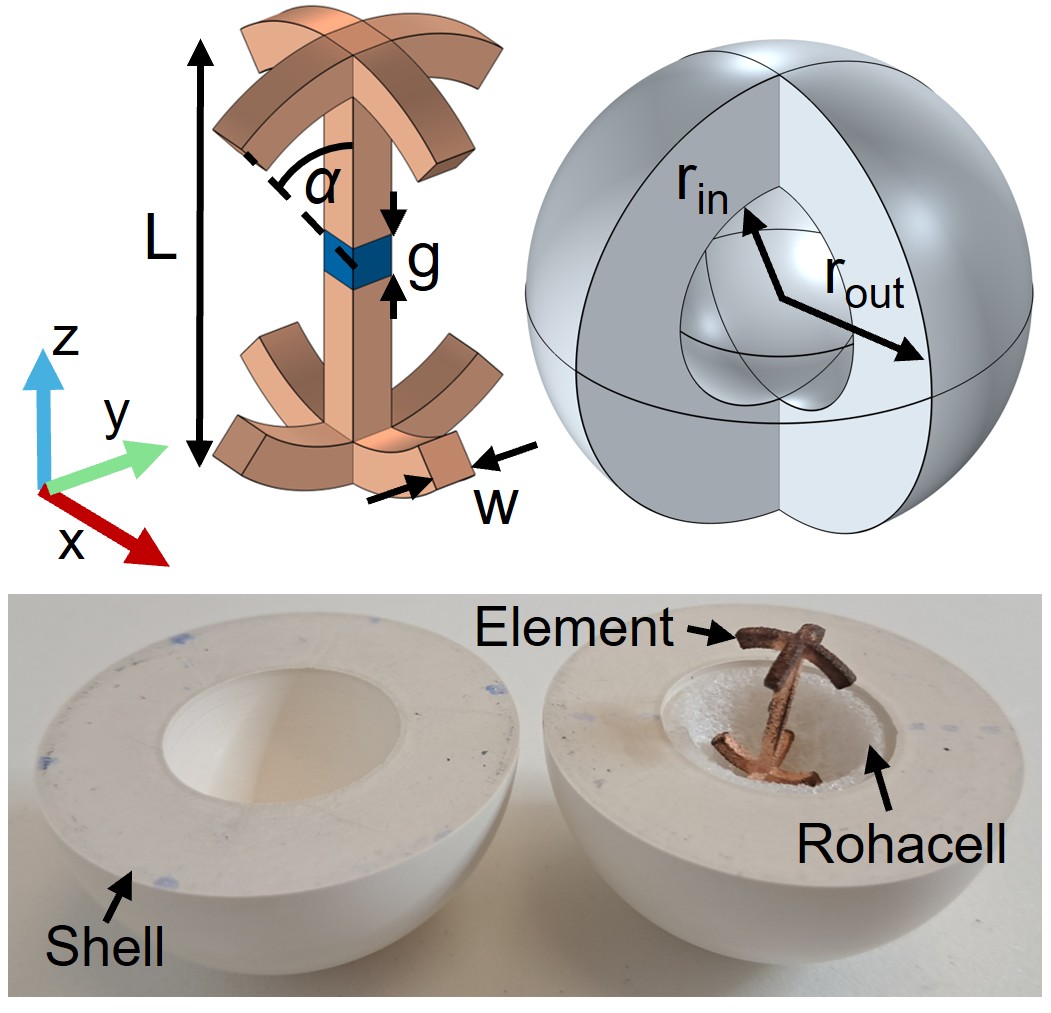}
\caption{Schematic showing the modelled metal element (top left) and dielectric shell (top right). The element's length L is aligned to the z-axis. The highlighted midpoint region indicates the location of where the gap / PIN diode is if present and has size given by g. $\alpha$ is the angle from the centre to the end of an arm and w is the width. r$_{\text{out}}$ is the radius of the shell and r$_{\text{in}}$ is the radius of the air core. A section of the shell is missing in this visual representation so that the core is visible. The photo (bottom) shows a fabricated sample consisting of Avient Preperm PPE500 shell, the copper electroplated metal element and a Rohacell 31 IG shell.}
\label{fig:dimensions}
\end{figure}

In this current study, we obtain the Kerker effect using a dielectric shell surrounding a metal element (\figref{fig:dimensions}). For an incident, z-polarised wave in the x-y plane, excitation of the dielectric shell provides a magnetic dipole moment perpendicular to the z-axis. With the orthogonal electric dipole contribution coming from the metal element aligned parallel to z and placed in the air filled core. An electric dipole can be obtained from a flat design on a printed circuit board. However, the magnitude will vary over angle of incidence in the x-y plane, deviating away from the requirements of the Kerker effect. Due to the 3-dimensional design of the metal element, the magnitude of the electric dipole is more stable over angle of incidence in the x-y plane than a printed circuit board design. Therefore the Kerker minimum arising from placing the metal element at the centre of the dielectric shell also deviates less. The arms at the ends of the metal element tuned its resonant frequency while allowing it to fit within the air core. By electrically disconnecting the two halves of the metal element at its mid point and placing a PIN diode across it, we can then effectively switch the electric dipole. At our design frequency, an `off' state is achieved when no bias is applied to the PIN diode, the geometry is effectively broken in half; for the `on' state, a bias is applied so the PIN diode then behaves as a resistor and an electric dipole is re-obtained. Switching the PIN diode bias allows for a backscatter minimum to be turned on and off.

\section{Method}
\label{sec:method}
To make the metal element, the base structure was first resin 3D printed. Coats of acrylic and then nickel conductive paint were then applied via spray can. Finally the structure was copper electroplated to achieve a full metal coating with average dimensions as given in \figref{fig:dimensions}. Where L = 14.435 mm, w = 1.380 mm and $\alpha$ = 40.7°. The outer radius of the arms is L/2. A photo of a fabricated sample can also be seen in \figref{fig:dimensions}. Some elements are deliberately metal coated with a gap (g$\simeq$1mm) at their midpoint in order to electrically disconnect the two ends of the structure. After electroplating, an Infineon BAR50-02V RF PIN diode and two copper wires were stuck onto these metal elements using silver DAG conductive paint. The PIN diode was placed spanning the gap, with one copper wire either end, wrapped around the element for ease of fabrication. The copper wires allow for biasing the PIN diode. In the results section, we discuss the response of the element when the gap and the PIN diode is present. 

The shell geometry was milled as two hemispheres from Avient Preperm PPE500 and has the dimensions given in \figref{fig:dimensions} where r$_{\text{in}}$ = 9.060 mm and r$_{\text{out}}$ = 18.500 mm. Two holes of radius 0.15 mm were drilled into one of the hemispheres so the copper wires could be threaded through. Two hemispheres of Rohacell 31 IG were milled and used to hold the metal element in the middle of the air core, so that the electric and magnetic dipole moments would be co-located, with the element aligned in z.

The scattering of the designs has been characterized using standard radar cross-section (RCS) techniques in an anechoic chamber \cite{Balanis2008}. Two Flann Microwave dual polarised horns (Model DP240-AB) were placed next to each other and connected to an Antritsu Shockline MS46122B 20 GHz Vector Network Analyzer to obtain quasi-monostatic (backscatter) RCS. For different incident angles, the sample being measured was rotated about its z-axis. To obtain bistatic RCS, one of the horns was rotated around the sample in the x-y plane, while the other remained stationary. In these RCS measurements, the electric field is parallel to the z-axis. 

For the biasing measurement, the copper wires attached to the metal element were connected to a power supply external to the anechoic chamber. A forward bias of 80 mA and 1.1 V was applied for the on state and zero bias applied for the off state. The bias wires were perpendicular to the electric field polarisation. The experimental data has been time-gated during post processing to filter out unwanted scattering events. 

Modelling was done using a finite element method solver \cite{COMSOLv6.4}. The models use the same dimensions as given for \figref{fig:dimensions}. The metal element was modelled as perfect electrical conductor. The dielectric shell was modelled with relative permittivity $\epsilon_r$ = 4.9 and a loss tangent of $\tan \delta$ = 0.0009 to represent the Avient Preperm PPE500. The Rohacell 31 IG hemispheres were modelled with relative permittivity $\epsilon_r$ = 1.06 and a loss tangent of $\tan \delta$ = 0.0011. The modelling allowed us to understand the influence of the holes drilled into the dielectric. The holes had negligible effect. The inclusion of the Rohacell in the model showed it shifted the Kerker minimum by $-$0.04 GHz. RCS data for both experiment and model has been plotted in decibels with the unit dBsm meaning decibels relative to one square metre.

\section{Results}
\label{sec:results}

\begin{figure*}[h!]{}
\centering
\includegraphics[width=160mm]{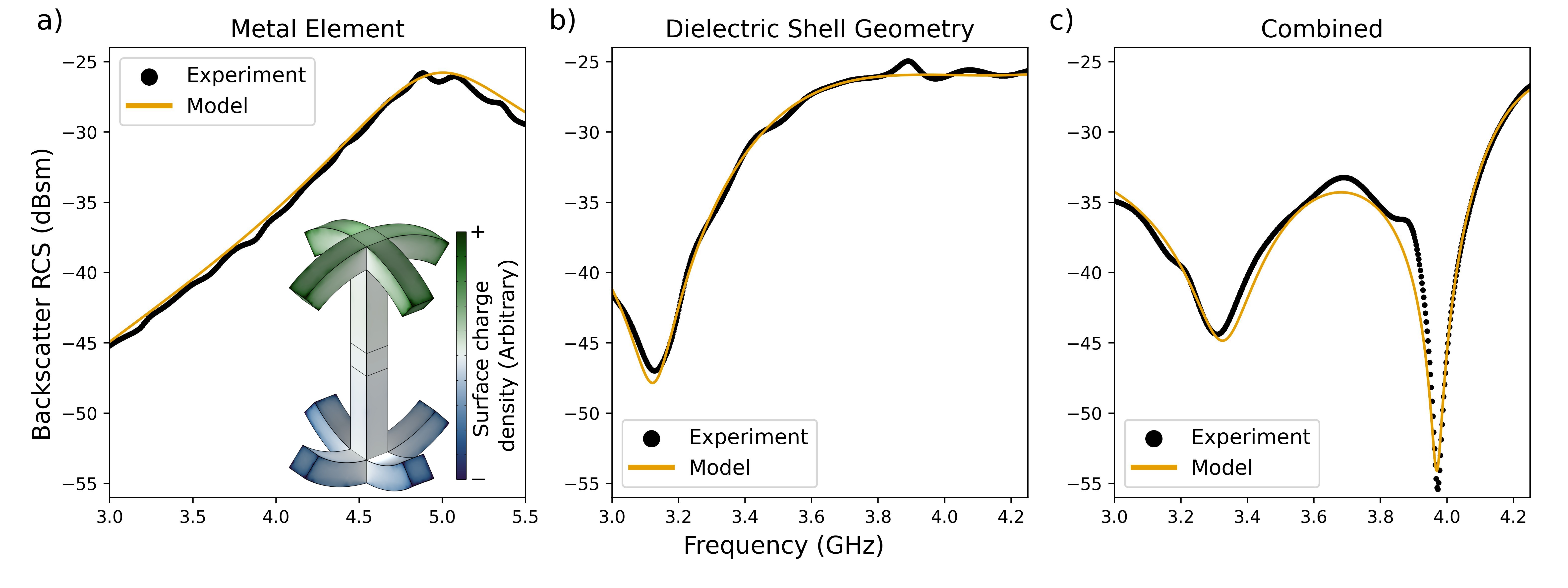}
\caption{Backscatter RCS of the geometries in \figref{fig:visual} vs frequency from both modelling and experiment. Experiment is quasi-monostatic RCS. Results are shown in dBsm. a) The metal element with an inset showing the surface charge density at 5.00 GHz normalized to maximum magnitude value and saturated. b) The dielectric shell geometry. c) The metal element located at the centre of the dielectric shell.}
\label{fig:structures}
\end{figure*}

Experimental quasi-monostatic (backscatter) RCS results are compared to modelling for the metal element, dielectric shell and then both combined in \figref{fig:structures} (plots a, b and c respectively). In \figref{fig:structures}a we see the metal element's fundamental resonance at 5.00 GHz, that will provide an electric dipole contribution to the Kerker effect in the combined geometry. The inset of \figref{fig:structures}a shows the surface charge density of the metal element at the fundamental resonance.

\begin{figure*}[h!]{}
\centering
\includegraphics[width=160mm]{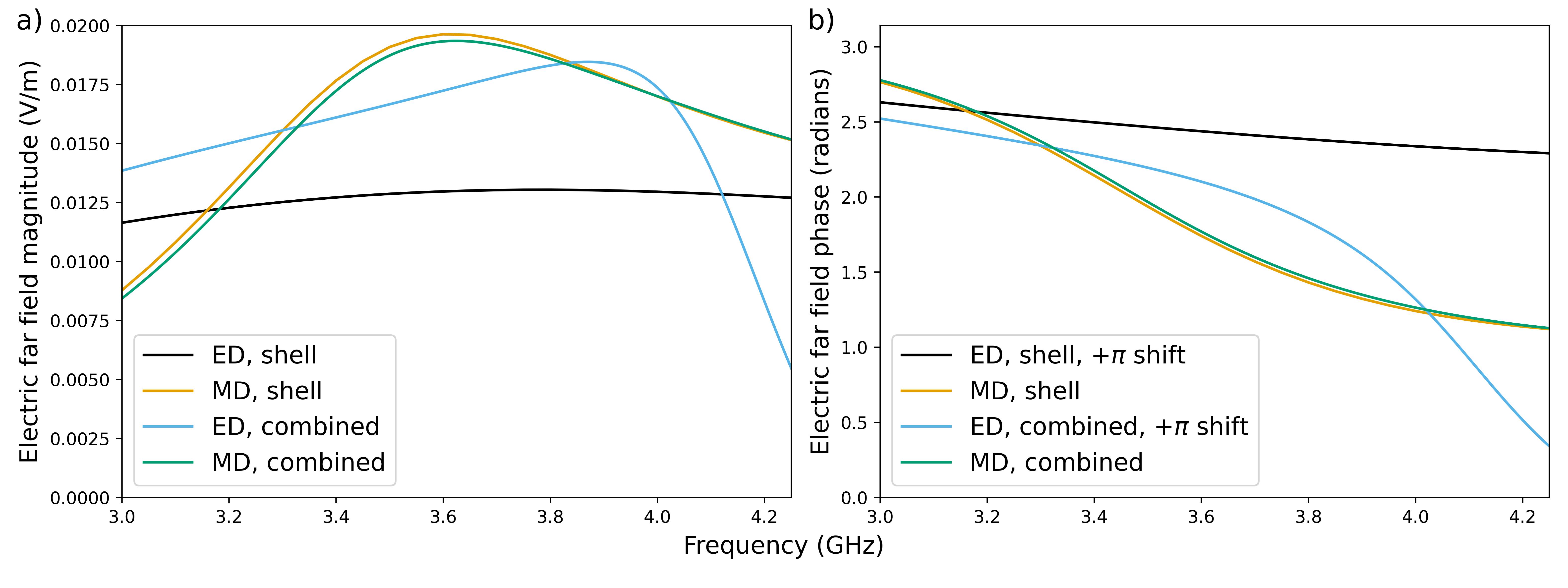}
\caption{Dipolar contributions to the backscattered electric far field's a) magnitude and b) phase obtained from modelling. This is for the dielectric shell and the combined geometry modelled in \figref{fig:structures} b and c respectively. The phase of the electric far field from the electric dipole contribution has been shifted by $+\pi$ to make clearer where it is $+\pi$ out-of-phase with the magnetic dipole.}
\label{fig:decomp}
\end{figure*}

To explain the backscatter RCS of the dielectric shell in \figref{fig:structures}b, we make use of a multipole decomposition that is obtained in the supplementary material. Through this the electric far field's magnitude and phase in the backscatter direction for the electric (ED) and magnetic (MD) dipoles are obtained from modelling. In \figref{fig:decomp}a the magnitude of the dipole contributions is plotted. The dielectric shell supports a magnetic dipole resonance at 3.60 GHz and an electric dipole resonance at 3.77 GHz that intersect in magnitude at 3.16 GHz. In \figref{fig:decomp}b it can be seen that the electric and magnetic dipole contributions of the dielectric shell are $\pi$ out-of-phase at 3.16 GHz. Therefore the backscattered electric far fields destructively interfere, resulting in the minimum at 3.13 GHz in \figref{fig:structures}b. While this is the Kerker effect, it is not of interest in the current work because it is not practically switchable. Our multipole decomposition does not obtain higher order contributions and the intersects in magnitude and phase do not occur at exactly the same frequency, hence the mismatch between frequencies of the intersects and the associated minimum. 


The experimental and modelled results for the combined geometry (no PIN diode), as illustrated in \figref{fig:dimensions}, are shown in \figref{fig:structures}c. There are two minima, at 3.31 GHz and 3.97 GHz, both associated with the Kerker effect. To help understand this plot, we again use multipole decomposition. In \figref{fig:decomp}a it can be seen that the addition of the metal element to create the combined structure changes the electric dipole moment contribution from that of just the dielectric shell. The combined structure has equal electric far field magnitudes from the dipoles at 3.33 GHz, 3.83 GHz and 4.02 GHz. There are minima from the Kerker effect in \figref{fig:structures}c at 3.31 GHz and 3.97 GHz, associated with the first and last intersect of magnitudes in \figref{fig:decomp}a. There is no Kerker effect minimum associated with the 3.83 GHz magnitude intersect and we look to the phase of the electric far field to see why. In \figref{fig:decomp}b it can be seen that the electric and magnetic dipole contributions are $\pi$ out-of-phase at 3.32 GHz and 4.02 GHz and so destructively interfere to produce minima. Meanwhile at 3.83 GHz there is 2.79 radians phase difference between the dipoles' contributions and so the interference does not result in a minimum. The influence of the dielectric surrounding the metal element shifts its resonance down from 5.00 GHz, hence why the associated minimum appears at a lower frequency than might be initially expected. 

\begin{figure}[h!]{}
\centering
\includegraphics[width=80mm]{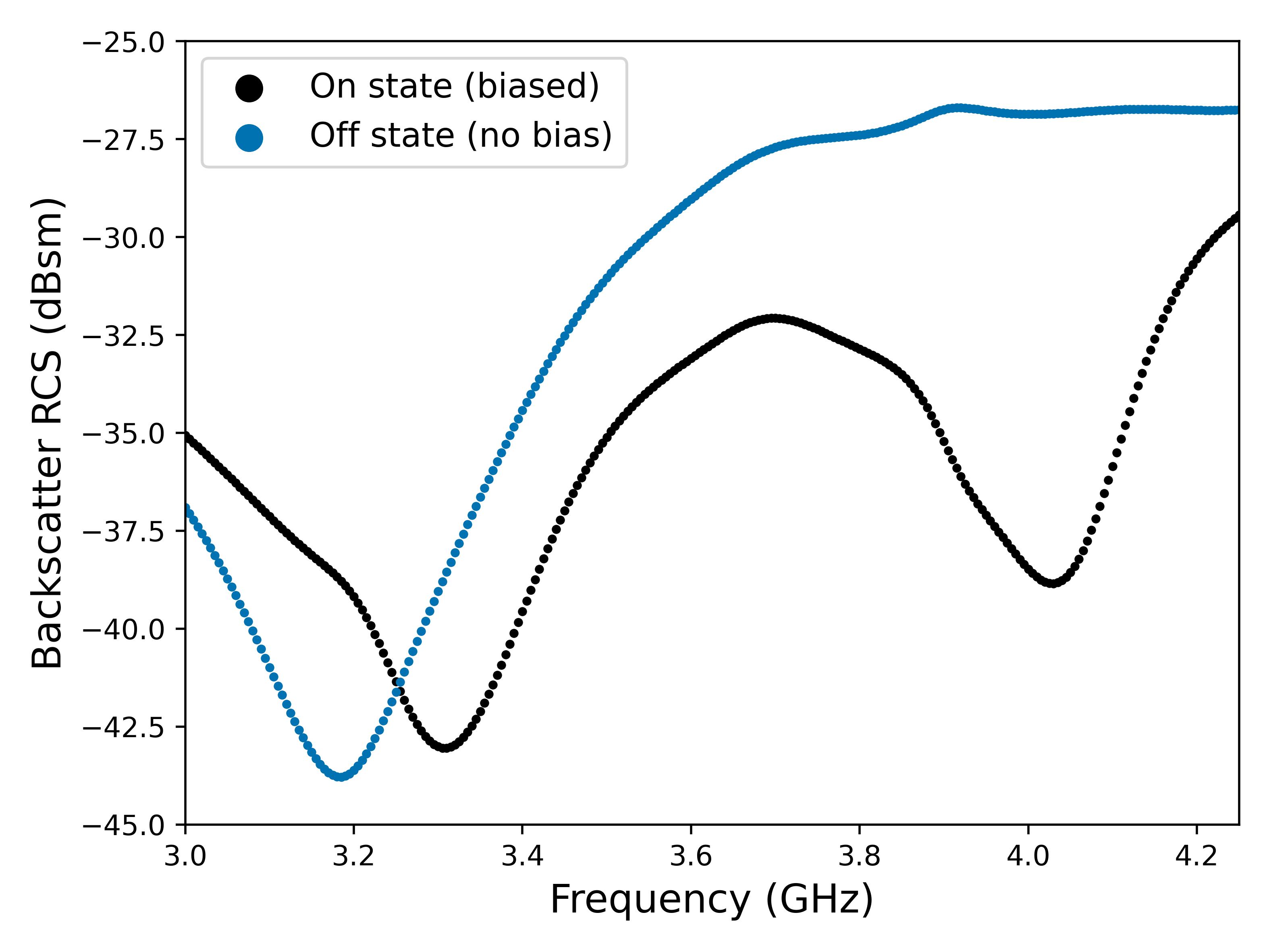}
\caption{Experimental quasi-monostatic (backscatter) RCS vs frequency for the on and off states of a metal element with PIN diode that is located at the centre of a dielectric shell. The on (off) state is with (without) a bias applied. Results are shown in dBsm.}
\label{fig:bias}
\end{figure}

Having obtained a structure exhibiting the Kerker effect, we then experimentally demonstrated its ability to be switched as observed in terms of the backscatter RCS plotted in \figref{fig:bias}. The switching was achieved using the combined geometry with the PIN diode inserted. With a bias applied, the PIN diode acts as to connect the two halves of the metal element, resulting in a minimum in the RCS associated with the Kerker effect at 4.03 GHz. With no bias applied, the metal element's electric dipole is not resonant at these frequencies and so the corresponding Kerker effect is switched off. 

To apply a bias, copper wires connected the diode to a power supply. To minimise their scattering contribution, these were aligned perpendicular to the incident electric field. 
It is not possible to accurately represent the attachment of PIN diode and wires in modelling. Therefore, we refer back to the case of no PIN diode and wires in \figref{fig:structures} for comparison. The off state is alike to the dielectric shell geometry but with variation due to the presence of the wires and resonances of the metal element located at higher frequencies. The on state is alike to the combined geometry but there is variation due to the wires and resistance of the PIN diode. This experiment using a PIN diode is a clear demonstration of this system displaying a switchable Kerker effect at $\sim$4 GHz.


\begin{figure*}[h!]{}
\centering
\includegraphics[width=160mm]{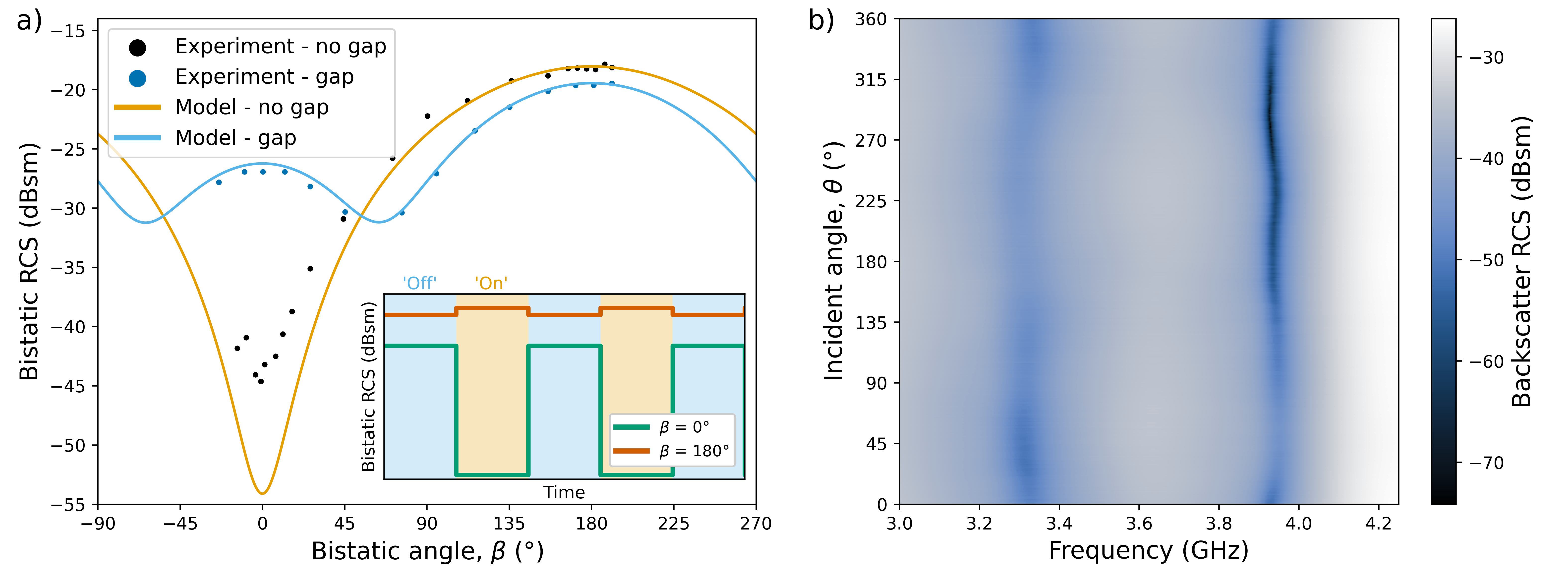}
\caption{a) Bistatic RCS vs bistatic angle ($\beta$) at 3.95 GHz from both modelling and experiment. All data is for the metal element at the centre of the dielectric shell. The label specifying gap or not refers to if one is present at the centre of the metal element. The inset shows a visual schematic of the modelled bistatic RCS at $\beta=$ 0° (backscatter direction) and 180° (forward scatter direction) in the scenario of switching between `on' and `off' states over time.  b) Colour plot of quasi-monostatic (backscatter) RCS as a function of the incident angle ($\theta$) and frequency for a metal element (no gap) at the centre of the dielectric shell. Results are shown in dBsm.}
\label{fig:bistatic}
\end{figure*}

To demonstrate that the contrast is greatest in backscatter, we measured the bistatic RCS of two versions of the combined geometry using the same dielectric shell but different metal elements. One with a gap (equivalent to bias `off') and another without (equivalent to bias `on'). The bistatic RCS at 3.95 GHz for these two variants is shown in \figref{fig:bistatic}a. 3.95 GHz being the frequency of the minimum arising from the destructive interference of the metal element's dipole moment with that of the sphere. The angle $\beta$ is between the two horns in the x-y plane, where 0° and 180° are the backward and forward scatter directions respectively. At $\beta$= 0° the model shows the magnitudes of the two states to be $-$26.2 dBsm and $-$54.1 dBsm. By 90° this has reduced to $-$27.8 dBsm and $-$23.8 dBsm. Finally by 180° it is $-$19.5 dBsm and $-$18.0 dBsm. In other words the difference between the states is greatest in backscatter direction and the contrast considerably decreases over other bistatic angles. The inset of \figref{fig:bistatic}a provides a schematic representation of what amplitude switching between the two states might look like: at 0° there is significant contrast in RCS while at 180° it is weaker. The bistatic angle label refers to the angle between the incident and received waves, and so can be considered the angle of an observer.  Therefore this data supports the idea of using the Kerker effect to increase security of communication against unwanted observers. Although the no gap variant is the same sample as in \figref{fig:structures}c, there is a minor difference in frequency and magnitude of this minimum at backscatter (0°) arising from uncertainty in the metal element's orientation and placement with respect to the centre of the shell.

\figref{fig:bistatic}b shows the angular stability of the Kerker effect via the backscatter RCS measurements of the combined structure (the same as measured in \figref{fig:structures}c) for different angles of incidence ($\theta$) in the x-y plane. The 3.94 GHz minimum remains present across all incident angles with a magnitude remaining below $-$45 dBsm. The metal element is sensitive to its placement within the air core and small deviation from the centre results in the minor variation in frequency and magnitude as the structure is rotated. \figref{fig:bistatic} shows the frequency varying from 3.93 to 3.95 GHz. These results demonstrate that this design has a stable Kerker effect present for all incident angles in the x-y plane.

\section{Outlook}
\label{Outlook}
This work demonstrates the potential for the Kerker effect to modulate a signal only in the backscatter direction. The Kerker effect is the destructive interference in the backscatter radiation obtained from the overlapping fields of an electric and magnetic dipole. To obtain this, our design consisted of a dielectric shell with a metal element at its centre, providing a magnetic and a co-located electric dipole respectively. The introduction of a PIN diode to the middle of the metal element allows for the shifting of the electric dipole moment away from the frequency of interest, switching off the minimum in backscatter radiation. This was experimentally demonstrated, showing that the resulting change of magnitude is greatest at backscatter. We also experimentally showed that due to the symmetry of the design, the minimum remains for any incident angle in the x-y plane. We hope that this design may find use in secure communications in order to reduce the possibility of unwanted observers of the scattering modulation at directions other than backscatter. Future work could explore reducing the difference in bistatic RCS magnitude between on and off states at non-backscatter angles. Future designs could also work towards maintaining switchable Kerker effect's angular stability for any incident wave, not just for those in one plane, such as in previous work for fixed (non-switchable) scatterers \cite{Powell2024}. 

\section*{Acknowledgements}
KJC acknowledges financial support by an ICASE from the Engineering and Physical Sciences Research Council (EPSRC) and QinetiQ (Grant number EP/X524906/1, voucher number 220181). CMH acknowledges finanical support from EPSRC via the Exeter University Physics DTP. For the purpose of open access, the author has applied a Creative Commons Attribution (CC BY) licence to any Author Accepted Manuscript version arising from this submission.

The authors have no conflicts of interest to disclose.

\section{Author Contributions}
\textbf{Kieran Cowan:} Conceptualization, Data Curation, Formal Analysis, Investigation, Methodology, Validation, Visualization, Writing - original draft. \textbf{Calvin Hooper:} Formal analysis. \textbf{Camergon Gallagher:} Investigation, Methodology. \textbf{Simon Berry:} Conceptualization, Project administration, Resources, Supervision, Writing - review \& editing.  \textbf{Alastair Hibbins:} Conceptualization, Funding acquisition, Project administration, Resources, Supervision, Writing - review \& editing.  \textbf{Alex Powell:} Conceptualization, Funding acquisition, Project administration, Resources, Supervision, Writing - review \& editing.

\printbibliography

\end{document}


\title{Supplementary material for Switchable selective backscatter modulation via the Kerker effect} 
\author{Kieran J. Cowan,$^{1, a)}$ Calvin M. Hooper,$^1$ Cameron P. Gallagher,$^1$ Simon J. Berry,$^2$ \and Alastair P. Hibbins,$^1$ and Alexander W. Powell,$^1$}

\date{$^1$Department of Physics and Astronomy, Centre for Metamaterial Research and Innovation, University of Exeter, Stocker Road, Exeter EX4 4QL, UK\\
    $^2$QinetiQ, Farnborough, Hampshire, GU14 0LX, UK\\
    $^{a)}$kc485@exeter.ac.uk\\[2ex]}
    
\onecolumn	
\maketitle 

In this supplementary we explicitly derive a well understood result that obtains the dipolar contributions to the electric far field \cite{Jackson}. While simpler equations can be obtained that are considered easier to work with, we found them incompatible with our work due to the structure being comparable in size to the wavelength and limitations in modelling \cite{Alaee2018, Evlyukhin2016}. Therefore we work through it ourselves here to be able to obtain the electric and magnetic dipole contributions to the electric far field using the output electric far field components from COMSOL \cite{COMSOLv6.4}.

The multipole decomposition of the electric ($\vec{E}$) and magnetic ($\vec{H}$) fields at point $\vec{r}$ and time, t, is given by

\[\vec{E}(\vec{r}, t) = \Re \left( \sum_{\ell=0}^\infty\sum_{m=-\ell}^\ell\left[a_{\ell m}^{(M)}h_\ell^{(1)}(kr)\vec{X}_{\ell m}(\theta, \phi) + \frac{iZ_0}{k}a_{\ell m}^{(E)}\nabla\times(h_\ell^{(1)}(kr)\vec{X}_{\ell m}(\theta, \phi))\right]e^{(-i\omega t)}\right)\]
and
\[\vec{H}(\vec{r}, t) = \Re \left( \sum_{\ell=0}^\infty\sum_{m=-\ell}^\ell\left[a_{\ell m}^{(E)}h_\ell^{(1)}(kr)\vec{X}_{\ell m}(\theta, \phi) - \frac{i}{kZ_0}a_{\ell m}^{(M)}\nabla\times(h_\ell^{(1)}(kr)\vec{X}_{\ell m}(\theta, \phi))\right]e^{(-i\omega t)}\right),\]
where $\vec{X}_{\ell m}(\theta, \phi)$ are normalised vector spherical harmonics, $h_\ell^{(1)}(kr)$ and $h_\ell^{(2)}(kr)$ are spherical Hankel functions, $a_{\ell m}^{(E)}$ and $a_{\ell m}^{(M)}$ are the electric and magnetic multipole coefficients respectively, $k$ is the wavenumber, $\omega =k/c$, $Z_0$ is the impedance of free space and $r^2=|\vec{r}|^2$ \cite{Jackson}.

The relevant multipole moments are \(h_{l}^{(1)}\vec{X}_{lm}\), and \(\frac{1}{k}\nabla \times \left( h_{l}^{(1)}\vec{X}_{lm} \right)\). However, we are interested in their far field behaviour, so let us consider their asymptotics. Since \(h_{l}^{(1)}\vec{X}_{lm}\) is only multiplied by a scalar function of \(r\), the asymptotic field profile can immediately be identified as simply \(\vec{X}_{lm}\). For \(\frac{1}{k}\nabla \times \left( h_{l}^{(1)}\vec{X}_{lm} \right)\), however, we must be a little more careful:

\[\nabla \times \left( h_{l}^{(1)}(kr)\vec{X}_{lm}(\theta,\phi) \right) = h_{l}^{(1)}(kr)\left( \nabla \times \vec{X}_{lm}(\theta,\phi) \right) + \left( \nabla h_{l}^{(1)}(kr) \right) \times \left( \vec{X}_{lm}(\theta,\phi) \right).\]

We may then consider the scaling of each of these terms with \(r\) and \(k\):

\[\left| h_{l}^{(1)}(kr) \right|\sim\frac{1}{kr},\]
    
\[\left| \nabla h_{l}^{(1)}(kr) \right|\sim\frac{1}{r},\]

\[\left| \vec{X}_{lm}(\theta,\phi) \right|\sim 1,\]

\[\left| \nabla \times \vec{X}_{lm}(\theta,\phi) \right|\sim\frac{1}{r}.\]

Hence, the second term dominates, decaying as \(\frac{1}{r}\) rather than \(\frac{1}{kr^{2}}\). Noting that \(\left( \nabla h_{l}^{(1)}(kr) \right) \parallel \boldsymbol{\hat{r}}\), we then find our two far-field profiles for a given \(l\) and \(m\) to be \(\vec{X}_{lm}\) and \(\boldsymbol{\hat{r}} \times \vec{X}_{lm}\). One can immediately note that these are locally orthogonal.

We are then left considering the form of \(\vec{X}_{lm}\). In this notation, we find \cite{Jackson}:

\[\vec{X}_{lm}(\theta,\phi) \propto \vec{r} \times \nabla Y_{lm}(\theta,\phi),\]
where

\[Y_{lm}(\theta,\phi) \propto e^{im\phi}P_{l}^{m}\left( \cos(\theta) \right).\]

Thus, we begin by working out the scalar spherical harmonics for a few of the lowest angular momentum states:

\[P_{0}^{0}\left( \cos(\theta) \right) \propto 1,\]

\[P_{1}^{- 1}\left( \cos(\theta) \right) \propto \sin(\theta),\]

\[P_{1}^{0}\left( \cos(\theta) \right) \propto \cos(\theta),\]

\[P_{1}^{1}\left( \cos(\theta) \right) \propto \sin(\theta),\]
which allows us to retrieve the scalar spherical harmonics.

\[Y_{0}^{0}(\theta,\phi) \propto 1,\]

\[Y_{1}^{- 1}(\theta,\phi) \propto e^{- i\phi}\sin(\theta),\]

\[Y_{1}^{0}(\theta,\phi) \propto \cos(\theta),\]

\[Y_{1}^{1}(\theta,\phi) \propto e^{i\phi}\sin(\theta).\]

For convenience, we then define:

\[Y_{1,x}(\theta,\phi) = \cos(\phi)\sin(\theta) \propto AY_{1}^{- 1}(\theta,\phi) + BY_{1}^{1}(\theta,\phi),\]

\[Y_{1,y}(\theta,\phi) = \sin(\phi)\sin(\theta) \propto AY_{1}^{- 1}(\theta,\phi) - BY_{1}^{1}(\theta,\phi),\]

\[Y_{1,z}(\theta,\phi) = \cos(\theta) \propto Y_{1}^{0}(\theta,\phi),\]
noting that

\[Y_{1,q} = \boldsymbol{\hat{r}} \cdot \boldsymbol{\hat{q}}.\]

Orthogonality of \(Y_{1,x}\) and \(Y_{1,y}\) is simple to confirm by noting that an integral over their product in \(\phi\) is z for all \(\theta\).

Note that \(Y_{0} = 1\) will differentiate to zero, and thus is not observed as a harmonic. we then define:

\[\vec{X}_{1,q} = \vec{r} \times \nabla\left( \boldsymbol{\hat{r}} \cdot \boldsymbol{\hat{q}} \right).\]

Hopping over to summation convention to work out that gradient:

\[\left( \nabla\left( \boldsymbol{\hat{r}} \cdot \boldsymbol{\hat{q}} \right) \right)_{i} = \partial_{i}r_{j}q_{j}\left( r_{k}r_{k} \right)^{- \frac{1}{2}} = \left( r_{k}r_{k} \right)^{- \frac{1}{2}}\partial_{i}r_{j}q_{j} + r_{j}q_{j}\partial_{i}\left( r_{k}r_{k} \right)^{- \frac{1}{2}} = r^{- 1}\delta_{ij}q_{j} - \frac{1}{2}r_{j}q_{j}r^{- 3}\partial_{i}\left( r_{k}r_{k} \right)\] 
\[= r^{- 1}\left( \delta_{ij}q_{j} - r_{j}q_{j}r^{- 2}\left( r_{k}\partial_{i}r_{k} \right) \right) = r^{- 1}\left( q_{i} - r_{j}q_{j}r^{- 2}\left( r_{k}\delta_{ik} \right) \right) = r^{- 1}\left( q_{i} - r_{i}r_{j}q_{j}r^{- 2} \right).\]

And converting back out:

\[\nabla\left( \boldsymbol{\hat{r}} \cdot \boldsymbol{\hat{q}} \right) = \frac{1}{\left| \vec{r} \right|}\left( \boldsymbol{\hat{q}} - \frac{\vec{r}\left( \vec{r} \cdot \boldsymbol{\hat{q}} \right)}{\left| \vec{r} \right|^{2}} \right),\]

\[\vec{X}_{1,q} = \vec{r} \times \frac{1}{\left| \vec{r} \right|}\left( \boldsymbol{\hat{q}} - \frac{\vec{r}\left( \vec{r} \cdot \boldsymbol{\hat{q}} \right)}{\left| \vec{r} \right|^{2}} \right) = \boldsymbol{\hat{r}} \times \boldsymbol{\hat{q}}.\]

These modes can quickly be identified as the far field radiation patterns for electric and magnetic dipoles by comparison to literature \cite{Griffiths, Jackson}:

\begin{equation}
\label{eq: cross ED}
    \vec{E}_{ref,q}^{ED} = \boldsymbol{\hat{r}} \times \left( \boldsymbol{\hat{r}} \times \boldsymbol{\hat{q}} \right),
    \tag{S1}
\end{equation}
\begin{equation}
\label{eq: cross MD}
\vec{E}_{ref,q}^{MD} = \boldsymbol{\hat{r}} \times \boldsymbol{\hat{q}}.
    \tag{S2}
\end{equation}
Strictly speaking, the standard result is that the full multipole fields are orthogonal, so it is worth proving that their asymptotic field profiles are too. To this end, we will draw upon the standard orthogonality result for spherical harmonics that

\[\int_{}^{}{Y_{lm}^{*}Y_{l^{'}m^{'}}d\Omega} = A_{lm}\delta_{ll^{'}}\delta_{mm^{'}},\]

for some amplitude \(A_{lm}\) which depends on convention \cite{Jackson}. However, since we will manually calculate the normalisation factors for our spherical harmonics, we will assume henceforth that \((l,m) \neq \left( l^{'},m^{'} \right)\).

Let us first consider:

\[\int_{}^{}{\vec{X}_{lm}^{*} \cdot \vec{X}_{l^{'}m^{'}}d\Omega} = \int_{}^{}{\left( \boldsymbol{\hat{r}} \times \nabla Y_{lm} \right)^{*} \cdot \left( \boldsymbol{\hat{r}} \times \nabla Y_{l^{'}m^{'}} \right)d\Omega}\]
\[= \int_{}^{}{\left( \left( \boldsymbol{\hat{r}} \cdot \nabla Y_{l^{'}m^{'}} \right)\left( \boldsymbol{\hat{r}} \cdot \nabla Y_{lm} \right)^{*} - \left( \boldsymbol{\hat{r}} \cdot \boldsymbol{\hat{r}} \right)\left( \left( \nabla Y_{lm} \right)^{*} \cdot \left( \nabla Y_{l^{'}m^{'}} \right) \right) \right)d\Omega} \]
\[= - \int_{}^{}{\left( \left( \nabla Y_{lm} \right)^{*} \cdot \left( \nabla Y_{l^{'}m^{'}} \right) \right)d\Omega},\]

where the first equality applies the vector quadruple product, and the final equality is achieved by noting that \(Y_{lm}\) has no gradient in the \(r\) direction and that \(\boldsymbol{\hat{r}} \cdot \boldsymbol{\hat{r}} = 1\).

In suffix notation:

\[\left( \partial_{i}Y_{lm} \right)^{*}\left( \partial_{i}Y_{l^{'}m^{'}} \right) = \left( \partial_{i}Y_{lm}^{*}\left( \partial_{i}Y_{l^{'}m^{'}} \right) \right) - Y_{lm}^{*}\partial_{i}^{2}Y_{l^{'}m^{'}},\]

\[\left( \left( \nabla Y_{lm} \right)^{*} \cdot \left( \nabla Y_{l^{'}m^{'}} \right) \right) = \nabla \cdot \left( Y_{lm}^{*}\nabla Y_{l^{'}m^{'}} \right) - Y_{lm}^{*}\nabla^{2}Y_{l^{'}m^{'}}\]

Thus it follows that:

\[\int_{}^{}{\vec{X}_{lm}^{*} \cdot \vec{X}_{l^{'}m^{'}}d\Omega} = \int_{}^{}{\left( Y_{lm}^{*}\nabla^{2}Y_{l^{'}m^{'}} \right)d\Omega} - \int_{}^{}{\left( \nabla \cdot \left( Y_{lm}^{*}\nabla Y_{l^{'}m^{'}} \right) \right)d\Omega}.\]

Since \(Y_{lm}\nabla Y_{l^{'}m^{'}}\) is parallel to the surface of the sphere, and smooth everywhere, the generalised Stokes theorem applied to a closed surface implies that

\[\int_{}^{}{\left( \nabla \cdot \left( Y_{lm}^{*}\nabla Y_{l^{'}m^{'}} \right) \right)d\Omega} = 0.\]

Meanwhile, on a unit sphere, the vector spherical harmonics obey the relation \(Y_{lm}\nabla^{2}Y_{l^{'}m^{'}} = - \left( l(l + 1) \right)Y_{lm}Y_{l^{'}m^{'}}\), and hence by orthogonality of the spherical harmonics:

\[\int_{}^{}{\vec{X}_{lm}^{*} \cdot \vec{X}_{l^{'}m^{'}}d\Omega} = - \left( l(l + 1) \right)\int_{}^{}{\left( Y_{lm}^{*}Y_{l^{'}m^{'}} \right)d\Omega} = - \left( l(l + 1) \right)A_{lm}\delta_{ll^{'}}\delta_{mm^{'}}.\]

To handle the electric multipoles, we once again apply the scalar quadrupole product formula to show

\[\int_{}^{}{\left( \boldsymbol{\hat{r}} \times \vec{X}_{lm}^{*} \right) \cdot \left( \boldsymbol{\hat{r}} \times \vec{X}_{l^{'}m^{'}} \right)d\Omega} = - \int_{}^{}{\vec{X}_{lm}^{*} \cdot \vec{X}_{l^{'}m^{'}}d\Omega}.\]

Finally, to demonstrate orthogonality of the electric and magnetic moments, we consider:

\[\int_{}^{}{\vec{X}_{lm}^{*} \cdot \left( \boldsymbol{\hat{r}} \times \vec{X}_{l^{'}m^{'}} \right)d\Omega} = \int_{}^{}{\left( \boldsymbol{\hat{r}} \times \nabla Y_{lm}^{*} \right) \cdot \left( \boldsymbol{\hat{r}} \times \left( \boldsymbol{\hat{r}} \times \nabla Y_{l^{'}m^{'}} \right) \right)d\Omega}\]

And for a final time, we apply the scalar quadrupole product formula:

\[\int_{}^{}{\vec{X}_{lm}^{*} \cdot \left( \boldsymbol{\hat{r}} \times \vec{X}_{l^{'}m^{'}} \right)d\Omega} = - \int_{}^{}{\left( \nabla Y_{lm}^{*} \right) \cdot \left( \boldsymbol{\hat{r}} \times \nabla Y_{l^{'}m^{'}} \right)d\Omega} = \int_{}^{}{\boldsymbol{\hat{r}} \cdot \left( \left( \nabla Y_{lm}^{*} \right) \times \left( \nabla Y_{l^{'}m^{'}} \right) \right)d\Omega}.\]

Via suffix notation, once again:

\[\epsilon_{ijk}\left( \partial_{j}Y_{lm}^{*} \right)\left( \partial_{k}Y_{l^{'}m^{'}} \right) = \epsilon_{ijk}\left( \partial_{j}Y_{lm}^{*}\left( \partial_{k}Y_{l^{'}m^{'}} \right) \right) - \epsilon_{ijk}Y_{lm}^{*}\left( \partial_{j}\partial_{k}Y_{l^{'}m^{'}} \right) = \epsilon_{ijk}\left( \partial_{j}Y_{lm}^{*}\left( \partial_{k}Y_{l^{'}m^{'}} \right) \right),\]

where the latter step follows by noting that \(Y_{lm}^{*}\left( \partial_{j}\partial_{k}Y_{l^{'}m^{'}} \right)\) is symmetric in \(j\) and \(k\), but is contracted against the Levi-Civita tensor. Out of suffix notation:

\[\left( \nabla Y_{lm}^{*} \right) \times \left( \nabla Y_{l^{'}m^{'}} \right) = \nabla \times \left( Y_{lm}^{*}\nabla Y_{l^{'}m^{'}} \right).\]

Hence our integral rearranges into the form:

\[\int_{}^{}{\vec{X}_{lm}^{*} \cdot \left( \boldsymbol{\hat{r}} \times \vec{X}_{l^{'}m^{'}} \right)d\Omega} = \int_{}^{}{\boldsymbol{\hat{r}} \cdot \left( \nabla \times \left( Y_{lm}^{*}\nabla Y_{l^{'}m^{'}} \right) \right)d\Omega} = \int_{}^{}{\left( \nabla \times \left( Y_{lm}^{*}\nabla Y_{l^{'}m^{'}} \right) \right) \cdot d\vec{S}},\]

where the final integral is taken over the unit sphere. However, since \(Y_{lm}^{*}\nabla Y_{l^{'}m^{'}}\) is smooth everywhere on the surface of the sphere, Stokes theorem may be applied, specialising to a closed surface to show that:

\[\int_{}^{}{\left( \nabla \times \left( Y_{lm}^{*}\nabla Y_{l^{'}m^{'}} \right) \right) \cdot d\vec{S}} = 0,\]

and hence:

\[\int_{}^{}{\vec{X}_{lm}^{*} \cdot \left( \boldsymbol{\hat{r}} \times \vec{X}_{l^{'}m^{'}} \right)d\Omega} = 0.\]

Thus, orthogonality of the spherical harmonics imply orthogonality of the asymptotics of the vector spherical harmonics.

The total electric field can be written as a sum of vector spherical harmonics (\(VSH\)) as

\begin{equation}
    \label{eq: VSH sum}
    \vec{E}(\vec{r}) = \sum_{\beta \in VSH}C_\beta\vec{E}_\beta(\vec{r}).
    \tag{S3}
\end{equation}

We want to find the amplitudes \(C_\beta\). To this end, we want to do the integral of the dot product of the total field with some vector spherical harmonic denoted by n. We can substitute the total fields for \eqref{eq: VSH sum}. Due to orthogonality of the vector spherical harmonics, this is zero if \(\beta\ne n\), therefore we obtain
\begin{equation}
    \label{eq: integral}
    \int\vec{E}(\vec{r})\cdot\vec{E}_n(\vec{r})dA = C_n\int\vec{E}_n(\vec{r})\cdot\vec{E}_n(\vec{r})dA.
    \tag{S4}
\end{equation}
In this work, we are only interested in the electric and magnetic dipole, for which we have already obtained their far field radiation patterns in \eqref{eq: cross ED} and \eqref{eq: cross MD}. We can substitute \(\vec{E}_{ref, q}^{MD}\) in for \(\vec{E}_n(\vec{r})\) in the integral on the right hand side, take \(q\) to be x, y or z and integrate over the surface of a sphere. Converting to spherical coordinates and then integrating by substitution results in the integral being equal to \(\frac{8\pi R^2}{3}\), where \(R\) is the radius of the sphere.

Having shown this for a magnetic dipole we can show the same for an electric dipole by showing 
\[\vec{E}_{ref, q}^{MD} \cdot \vec{E}_{ref, q}^{MD} = \vec{E}_{ref, q}^{ED} \cdot \vec{E}_{ref, q}^{ED}.\]
Let \(\vec{b} = \boldsymbol{\hat{r}}\times \boldsymbol{\hat{q}}\) and then use the Lagrange identity to get
\[ \boldsymbol{\hat{r}} \times \vec{b} \cdot \boldsymbol{\hat{r}}\times \vec{b}  = |\boldsymbol{\hat{r}} \times\vec{b}|^2=(\boldsymbol{\hat{r}} \cdot \boldsymbol{\hat{r}})(\vec{b}\cdot\vec{b})-(\boldsymbol{\hat{r}}\cdot\vec{b})^2. \]
Can substitute \(\vec{b}\) back in and use \((\boldsymbol{\hat{r}} \cdot \boldsymbol{\hat{r}})=1\) so that the above simplifies down to
\[ =\boldsymbol{\hat{r}}\times\boldsymbol{\hat{q}} \cdot \boldsymbol{\hat{r}}\times\boldsymbol{\hat{q}} -\left( \boldsymbol{\hat{r}}\cdot(\boldsymbol{\hat{r}} \times \boldsymbol{\hat{q}} \right)^2 = \vec{E}_{ref, q}^{MD} \cdot \vec{E}_{ref, q}^{MD} -\left( \boldsymbol{\hat{r}}\cdot(\boldsymbol{\hat{r}} \times \boldsymbol{\hat{q}}) \right)^2. \]
then the scalar triple product can be used to show \(\left( \boldsymbol{\hat{r}}\cdot(\boldsymbol{\hat{r}} \times \boldsymbol{\hat{q}}) \right)^2=0\). 

Finally we can write the electric field given by an electric or magnetic dipole (\(\alpha D\) with \(\alpha \in E,M\)) aligned in direction q to be
\begin{equation}
    \label{eq: ED MD}
    \vec{E}^{\alpha D}(\vec{r}) = \sum_{q\in x,y,z}C_q^{\alpha D}\vec{E}_{ref, q}^{\alpha D}(\vec{r}),
    \tag{S5}
\end{equation}
where 
\begin{equation}
    \label{eq: Cq}
    C_q^{\alpha D} = \frac{3}{8\pi R^2}\int\vec{E}(\vec{r})\cdot \vec{E}_{ref, q}^{\alpha D}dA.
    \tag{S6}
\end{equation}
In our use, the surface integral is performed in COMSOL at the inner boundary of the perfectly matched layer with \(\vec{E}(\vec{r})\) being the electric far field. We are then only interested in the case of backward scattering, and so sub in \(\boldsymbol{\hat{r}}=-\boldsymbol{\hat{y}}\) (as we modelled a plane wave incident in the +y direction) for \ref{eq: cross ED} and \ref{eq: cross MD}. 

\printbibliography